# Developing a custom GPT based on Inquiry Based Learning for Physics Teachers

Dimitrios Gousopoulos, Panteion University of Social and Political Sciences, Greece

**Abstract**

Generative Artificial Intelligence (GenAI) has emerged as a valuable assistant in many fields such as marketing, finance, and project management and education. In education, many GenAI tools have been developed to aid teachers in preparing proper educational material and offering personalized learning to their students, tailored to their educational needs. In this paper we present a custom GPT (IBL Educator GPT) that is designed and developed based on Inquiry based Learning and offers physics teachers a framework in which they can interact with ChatGPT and design educational strategies. The utilization of the IBL Educator GPT has led to the improvement in teachers' perspectives regarding the adoption of artificial intelligence-based tools for personalizing teaching.

## 1. Introduction

Artificial intelligence and in particular Large Language Models (LLMs) such as ChatGPT are bringing rapid changes to science education. Through appropriate prompting, ChatGPT can improve personalized learning, students' problem-solving ability, and understanding of physics concepts [1,2,3]. Such tools can also assist in educational planning, introducing innovative teaching methods, and providing feedback to students based on their responses [3]. Despite all these advantages, there are several weaknesses that educators should consider seriously. Such is the inability of LLMs to manage content of increased difficulty and complexity, the lack of visual aids, as well as the need for human intervention to ensure the accuracy and correctness of the answers produced by these models [3]. The performance of LLMs varies at different academic levels. In particular, they produce more accurate answers to basic physics questions than university-level ones [2].

Regarding generative AI tools like ChatGPT and their use in the educational process, research has shown that on the one hand, it can contribute to the improvement of students' critical thinking [4], but on the other hand, they have been related to low academic performance and reduced autonomous learning [5]. In addition, generative AI tools can be used as learning assistants, and with their help, the teachers can design experiments that face students' alternative ideas about physical phenomena [6]. Despite their power, Generative AI tools must be utilised carefully and the answers they give must be checked in every case by the teacher.

One way to maximize the potential of LLMs like ChatGPT is through prompt engineering, which is a process of designing and improving the prompts provided in an LLM to craft better and more targeted responses from LLMs [7]. In an educational context, prompt engineering can improve the learner's learning experience by tailoring LLM responses to the specific learning needs of each class and student. In this way, critical thinking and personalized learning are further cultivated [8].

The custom GPT is the latest powerful feature of ChatGPT. Non-programmers can create and share their own GPTs ("chat bots"), allowing Science Educators to apply the capabilities of ChatGPT to create administrative assistants, online tutors, and more, to support their teaching environments.

## 2. Setup and task

The aim of this study is to create a custom GPT based on Inquiry Based Learning (IBL Educator GPT) [9], so that teachers can have an educational framework when interacting with ChatGPT and receive several perspectives concerning the subject they want to teach[10,11]. For each phase of IBL (orientation, conceptualization, investigation, conclusion and discussion), a prompt was crafted based on three pillars: Iterative refinement, role-playing and constraints and suitable context (Figure 1). Moreover, for each phase teachers can get 3 options to choose from and along with their evaluations so that they can have a basis upon which they can make a better decision (Figure 2,3).

> Act as an experienced educator. The users are teachers who will provide you with the subject they want to teach. Based on the selected subject, you should give them educational ideas and resources that are organized as follows:
> Orientation: is the process of stimulating curiosity about a topic and addressing a learning challenge through a problem statement
> Conceptualization: is the process of stating theory-based questions and/or hypotheses, generating research questions based on the stated problem, and generating hypotheses regarding the stated problem.
> Investigation: is the process of planning exploration or experimentation; collecting and analyzing data based on the experimental design or exploration
> Conclusion: is the process of drawing conclusions from the data by comparing inferences made based on data with hypotheses or research questions.
> Discussion: is the process of presenting findings of particular phases or the whole inquiry cycle by communicating with others and/or controlling the entire learning process or its phases by engaging in reflective activities, presenting outcomes of an inquiry phase or of the whole inquiry cycle to others (peers, teachers) and collecting feedback from them, describing, critiquing, evaluating and discussing the whole inquiry cycle or a specific phase.
> For each phase 1. propose 3 fundamentally different options, 2. Give the user 3 metrics to compare and contrast and 3. Evaluate these options using these metrics and before moving to the next phase, ask the user if they want to proceed to the next phase.

*Figure 1 The prompt behind the IBL Educator GPT*

**Phase 1: Orientation**

The goal here is to spark curiosity and identify a learning challenge related to Newton's Second Law.

**Option 1: Everyday Scenario**

- **Stimulus**: Ask students to imagine pushing a heavy shopping cart vs. a light one. How does it feel when trying to push them with the same force?
- **Problem Statement**: "Why does a heavier object require more force to accelerate than a lighter object?"

**Option 2: Sports Context**

- **Stimulus**: Show a video of different athletes (like a football player vs. a lightweight sprinter) running and being tackled.
- **Problem Statement**: "How do mass and force affect the acceleration of athletes in sports?"

**Option 3: Space Exploration**

- **Stimulus**: Present the scenario of a rocket launch. How does the force of the engines move such a massive object into space?
- **Problem Statement**: "How does force applied to a large mass, like a rocket, affect its acceleration compared to smaller objects?"

*Figure 2 Example of the GPT's proposed option regarding the 2nd Law of Newton*

## Metrics for Comparison

1. **Relevance**: How relatable is the scenario to students' daily lives or interests?
2. **Engagement**: How likely is it to captivate student attention?
3. **Clarity**: How well does the scenario introduce the concepts of force, mass, and acceleration?

## Evaluation of Options

- **Option 1 (Everyday Scenario)**: Highly **relevant** and **clear** since pushing carts is a common experience, though it might be less **engaging** than other options for older students.
- **Option 2 (Sports Context)**: Highly **engaging**, especially for students interested in sports, and somewhat **relevant**. However, it might require more explanation for full **clarity** regarding physics principles.
- **Option 3 (Space Exploration)**: Highly **engaging** for those interested in space but may be less **relevant** to everyday life. The **clarity** could be affected by the complexity of the rocket example.

*Figure 3 Example of the GPT's proposed metrics of comparison regarding the 2nd Law of Newton*

**3. Results**

For evaluating the developed GPT (you can have access here: https://chatgpt.com/g/g-GHyDSFa8h-ibl-educator)a pilot study was conducted. Fourteen science educators have tested its capabilities, and they filled in a questionnaire before the use of GPT and the same questionnaire after its use. The questions of the questionnaire were selected from published instruments in education and so they are characterized by validity and reliability [12]. The results showed that teachers' perspectives on the adoption of artificial intelligence-based tools for personalizing teaching improved. In particular, teachers have agreed that Artificial Intelligence can assist them in planning tasks such as preparing lessons and activities before class and improving teacher professional training. To observe the potential shift in teachers' beliefs concerning the adoption of AI tools in the classroom, the non-parametric Wilcoxon signed rank test applied. The test indicated that in post-test ranks were statistically significantly higher than pre-test ranks, $Z=-3.296$, $p<.001$. To conclude, GenAI tools can be used as teaching assistants that aid teachers gain time regarding their iterative daily tasks and present them several perspectives of each educational phase so that they can use this time to focus on designing and developing innovative teaching strategies.